\documentclass[useAMS,usenatbib]{mn2e}
\usepackage{epsfig}
\usepackage{graphicx}
\usepackage{amsmath}
\usepackage{amssymb}
\usepackage{bm,url,times}
\usepackage{float}

\def\Pm{\mbox{P}_M}
\def\Rm{\mbox{R}_M}

\def\Rey{\mbox{Re}}
\def\Urms{{u_{rms}}}
\def\Urms{{U_{rms}}}
\def\Brms{{B_{rms}}}

\def\apj{ApJ}

\def\mnras{MNRAS}
\def\aap{A\&A}
\def\apjl{ApJ}

\def\pre{PRE}

\def\apjs{ApJS}

\def\zhetp{JETP}

\newcommand{\EQ}{\begin{equation}}
\newcommand{\EN}{\end{equation}}
\newcommand{\EQA}{\begin{eqnarray}}
\newcommand{\ENA}{\end{eqnarray}}

\newcommand{\Eq}[1]{Eq.~(\ref{#1})}

\newcommand{\Eqss}[2]{Eqs.~(\ref{#1})--(\ref{#2})}

\newcommand{\Fig}[1]{Fig.~\ref{#1}}

\newcommand{\Figs}[2]{Figs.~\ref{#1} and \ref{#2}}

\newcommand{\bra}[1]{\langle #1\rangle}

{}
{}
{}

{}
{}
{}
{}
{}
{}
{}
{}
{}
\newcommand{\meanBB}{\overline{\mbox{\boldmath $B$}}{}}{}
{}
{}
{}
{}
{}
{}
{}
{}
{}
{}
{}
\newcommand{\meanUU}{\overline{\mbox{\boldmath $U$}}{}}{}

{}
{}
{}

{}

{}
{}
\newcommand{\emf}{{\cal E}}{}

\newcommand{\xxx}{\hat{\mbox{\boldmath $x$}} {}}
\newcommand{\yyy}{\hat{\mbox{\boldmath $y$}} {}}
\newcommand{\zzz}{\hat{\mbox{\boldmath $z$}} {}}

\newcommand{\bb}{\bm{b}}
\newcommand{\BB}{\bm{B}}

\newcommand{\UU}{\bm{U}}
\newcommand{\uu}{\bm{u}}

\newcommand{\JJ}{\mbox{\boldmath $J$} {}}

\newcommand{\AAA}{\mbox{\boldmath $A$} {}}

\newcommand{\nab}{\mbox{\boldmath $\nabla$} {}}

\newcommand{\DDD}{{\cal D} {}}

\def\la{\mathrel{\mathchoice {\vcenter{\offinterlineskip\halign{\hfil
$\displaystyle##$\hfil\cr<\cr\sim\cr}}}
{\vcenter{\offinterlineskip\halign{\hfil$\textstyle##$\hfil\cr<\cr\sim\cr}}}
{\vcenter{\offinterlineskip\halign{\hfil$\scriptstyle##$\hfil\cr<\cr\sim\cr}}}
{\vcenter{\offinterlineskip\halign{\hfil$\scriptscriptstyle##$\hfil\cr<\cr\sim\cr}}}}}
\def\ga{\mathrel{\mathchoice {\vcenter{\offinterlineskip\halign{\hfil
$\displaystyle##$\hfil\cr>\cr\sim\cr}}}
{\vcenter{\offinterlineskip\halign{\hfil$\textstyle##$\hfil\cr>\cr\sim\cr}}}
{\vcenter{\offinterlineskip\halign{\hfil$\scriptstyle##$\hfil\cr>\cr\sim\cr}}}
{\vcenter{\offinterlineskip\halign{\hfil$\scriptscriptstyle##$\hfil\cr>\cr\sim\cr}}}}}
\def\Pm{\mbox{\rm Pr}_M}
\def\Rm{R_{\rm m}}

\def\Rey{\mbox{\rm Re}}

\def\Brms{B_{\rm rms}}

\def\Urms{U_{\rm rms}}

\graphicspath{{./figs/}{./png/}}

\title[MRI as large scale dynamo]
{Large scale dynamo action precedes turbulence in shearing box simulations of the magnetorotational instability }

\author[]{Pallavi Bhat$^1$\thanks{
pbhat@princeton.edu}, Fatima Ebrahimi$^1$\thanks{febrahimi@princeton.edu} and Eric G. Blackman$^2$\thanks{blackman@pas.rochester.edu}\\
$^{1}$Department of Astrophysical Sciences and Princeton Plasma Physics Laboratory, Princeton University, Princeton, NJ 08543, USA\\
$^{2}$Department of Physics and Astronomy, University of Rochester, Rochester, NY14618, USA}
\begin{document}

\pagerange{\pageref{firstpage}--\pageref{lastpage}} \pubyear{2016}

\maketitle

\label{firstpage}

\begin{abstract}
We study the dynamo generation (exponential growth) of large scale (planar averaged)  fields 
in unstratified shearing box simulations of the magnetorotational instability (MRI).
In contrast to previous studies restricted to horizontal ($x$-$y$) averaging, we demonstrate the presence of large scale
fields when either horizontal {\it or} vertical ($y$-$z$) averaging is employed.
By computing planar averaged fields and power spectra, we find large scale dynamo action in the early MRI growth phase---a previously unidentified feature.
Fast growing horizontal low modes and fiducial vertical modes over 
a narrow range of wave numbers amplify these planar averaged fields in the MRI growth phase, before turbulence sets in. 
The large scale field growth requires linear fluctuations  but not nonlinear turbulence (as defined by mode-mode coupling) and grows as a direct global mode of the MRI. 
Only by \textit{vertical averaging},  can it be shown that the growth of 
horizontal low wavenumber MRI modes directly feed-back to the initial vertical field 
providing a clue as to  why  the large scale vertical field sustains against  turbulent diffusion in the 
saturation regime.  
We compute the terms in the planar averaged mean field equations to identify the individual contributions to large scale field growth for both vertical and horizontal averaging. 
The large scale fields obtained from such vertical averaging are found to compare well with global cylindrical simulations and quasilinear analytical analysis from a previous 
study by Ebrahimi \& Blackman. 
We discuss the potential implications of these new results for understanding large scale MRI dynamo saturation and turbulence.
\end{abstract}


\section{Introduction}

Magnetic fields have long been considered to play an important role in the generation of
turbulence in accretion discs via the magnetorotational instability (MRI) \citep{velikhov59, chandra60, BH91}.
Direct numerical simulations (local and global) of Keplerian discs now routinely demonstrate the sustanence of MRI-turbulence along with  large scale magnetic fields 
\citep{Axel95,Stone96,DSP10,KK11,Sorathia2012,GresselPessah2015,Shi2016}, where by "large scale"  we refer to mean fields that survive two-dimensional spatial  averaging over the simulation box.  These  planar averaged large scale fields also exhibit cycle periods 
on time scales of 10s of orbits, which have been semi-empirically modeled
with traditional mean field dynamo models.

While understanding how these large scale fields
are produced in  MRI unstable systems is of general interest 
for the connection to large scale dynamo theory,  their importance is greatly exacerbated by several recent realizations:
 (1) Stresses have been found to be time-correlated with the 
large scale field in stratified simulations of \citet{Guan2011} (an effect we also show here for unstratified simulations).
(2) Simulations also reveal that  the magnetic energy 
and Maxwell stresses are dominated by contributions from the lowest few wave numbers in the boxes.  
This suggests that the MRI in nature could involve significant contribution to large scale stresses 
rather than merely a "local" viscosity \citep{NB14}.  
(3) Stratified simulations show further that the ratio of the contribution to the stress from the mean 
fields to the fluctuations increase toward the lower density corona.
(4)  Astrophysical observations of jets and coronae in combination with these emerging lessons from  both local and nonlocal simulations
together suggest that transport in disks is  significantly non-local  for which large scale fields 
may be key \citep{BN15}.
All of this motivates the question: what mechanism is responsible for generating large scale
scale fields in the MRI simulations?

Previous studies of large scale dynamos in the context of shearing box 
simulations have typically focused on the secular long term evolution of large scale field 
in the presence of turbulence.
Modelling the cycle periods in the large scale fields is such an example  \citep{LesurOgil2008, Gressel2010, Simon2011, herault2011}.
An important ingredient in these models is the nonlinear mode
coupling which is considered responsible for generation of the field perpendicular to
the shear flow and rotation.
This role of nonlinear mode coupling is  also essential in dynamo models 
that describe the MRI turbulence in unstratified shearing boxes 
as a subcritical transition phenomenon \citet{rincon,Riolsetal2015}. 
Approaches based on turbulent helicity fluxes were used 
to model the MRI dynamo \citep{vishniac2009,ebrahimiprl}. More recently,
a magnetic shear-current effect based on fluctuating fields 
was proposed as the mechanism of large scale dynamo in MRI \citep{SB15}.

In contrast to these studies using mode-coupling/turbulence,  
\citet{EB16} (EB16 from here on) show from a single mode quasilinear analysis,
it is possible to obtain an EMF required for large scale dynamo action,
with the minimum requirement being that the perturbations need to be of non-axisymmetric nature.
This study did not focus on the secular cycle periods but on the initial exponential generation of vertically averaged mean fields
which emerges on the MRI growth time scale directly as a global mode for the size of the system under study. 
Linearized eigenfunctions were used to construct EMF in both the quasilinear analytical calculations 
and the single mode MRI (with a particular azimuthal and vertical mode number) simulations. 
The quasilinear analytical calculations were employed to explain the large-scale dynamo growth of 
radially alternating mean fields (averaged over height and azimuth) observed in the global cylindrical simulations.
Large-scale magnetic field generation due to a single global MRI mode resulting from its own EMF was first 
introduced in a global cylindrical model in \citet{ebrahimi2009}, but until now, has not been identified in the local shearing box simulations.  
In the present paper, we endeavour to clarify the nature of
large scale dynamo action in MRI simulations by analyzing  shearing box simulations
in a new way. To do so, we focus on the MRI in unstratified simulations with zero net flux,
which is the system with the simplest configuration.
\begin{table*}
\setlength{\tabcolsep}{3.5pt}
\begin{tabular}{|l|r|}
\hline
\hline
\textbf{Terms} & \textbf{Definition in the paper}  \\
\hline
Large scale field (or mean field) & planar averaged field,  $\meanBB$ (alternatively $\bra{\BB}$) \\
Large scale dynamo (or mean field dynamo) & exponential growth of 
any field component of $\meanBB$;\\
& or sustenance of any field component against  decay \\
EMF & $\bm{\emf}=\bra{\uu\times\bb}$, where $\uu=\UU-\meanUU$ and $\bb=\BB-\meanBB$	\\
\hline
\hline
\label{Tab}
\end{tabular}
\caption{Here we explicitly define the various terms used throughout the text to clarify
any ambiguities that may arise from different uses of these terms in other work.
We  define  large scale (or mean) fields as planar averaged fields and large  dynamo action
as the exponential growth of any component of a planar averaged field, (regardless of whether
the total mean field energy grows) or the sustenance of total magnetic energy of the large
scale field against decay.  This contrasts  other papers,  which sometimes view the large scale dynamo
from a perspective focusing on cycle periods.}
\end{table*}

\begin{figure}
\includegraphics[width=0.495\textwidth, height=0.32\textheight]{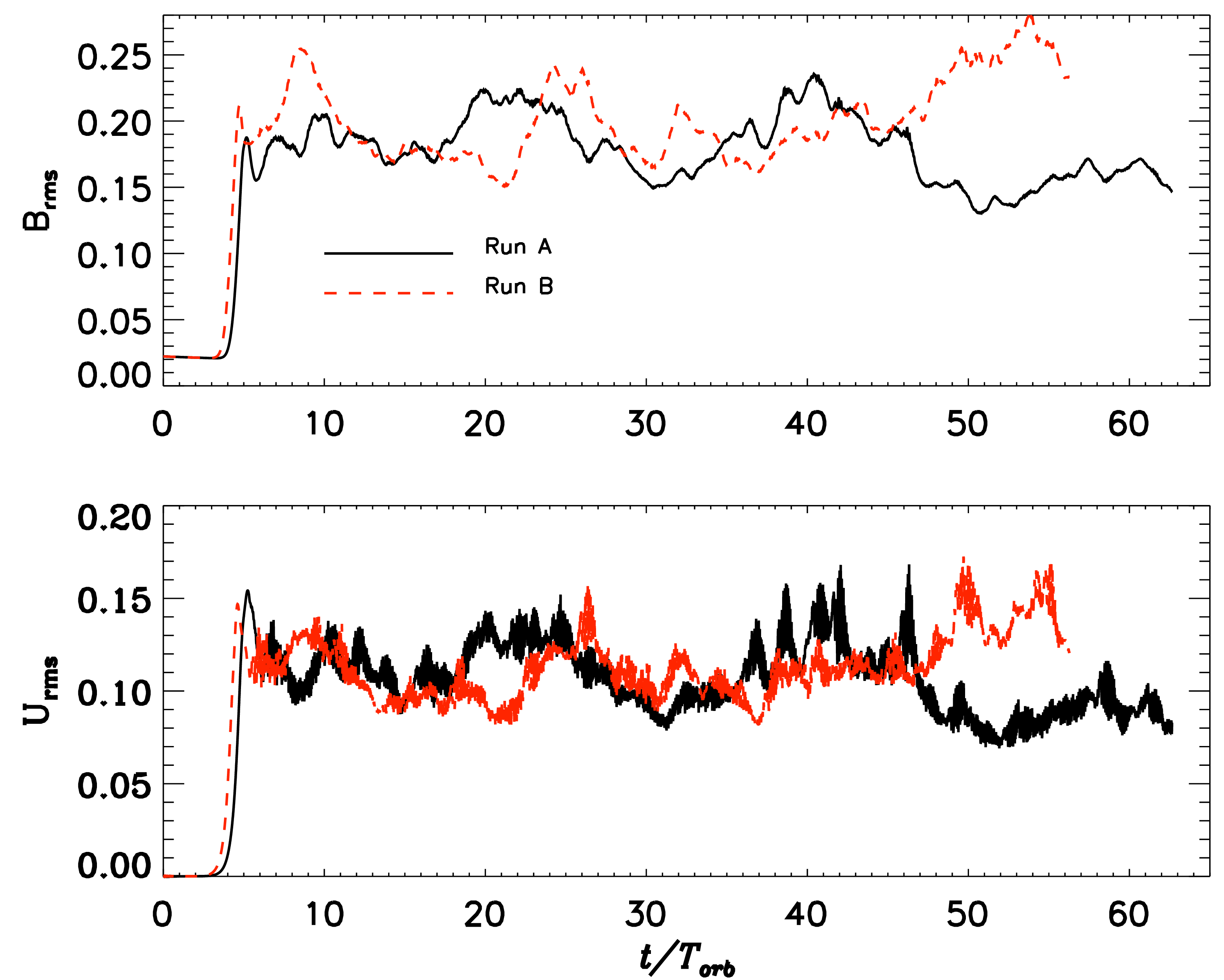}
\caption{In the upper panel, evolution of $\Brms$ and in the lower panel, evolution of $U_{rms}$
is shown. The solid black line is for Run~A and the dashed red is for Run~B. 
}
\label{timeplot}
\end{figure}

Although  the magnetic field evolution in the nonlinear saturation regime of the  MRI has been studied extensively in simulations, 
large scale dynamo action in the early MRI growth phase has heretofore not been explored. 
In contrast to previous MRI dynamo  shearing box simulation studies which focus on horizontal averaging
to obtain  mean fields, we consider separately horizontal and  vertical averaging and show the
presence of large scale fields in both approaches. We 
show that such large scale fields
emerge also during the early MRI growth phase.
Further, we perform spectral studies of the magnetic field evolution, 
not restricted to only vertical wavenumbers
but also in the horizontal $x$ and $y$ directions. This is of key importance
to identify the large scale dynamo action in the absence of turbulence. 
And to substantiate our findings, we evaluate the terms in the mean field equations
for a given type of averaging and show which terms are responsible for the dynamo action
in both early growth phase and the turbulent nonlinear regime.

In Table \ref{Tab}, we clarify the terms we use throughout the paper. As mentioned above, we
define large scale or mean fields as those which survive any planar averaging. 
Such averaged fields  may or may not have large gradients in the remaining 
unaveraged direction but still classify as large scale by our definition. 
Similarly, the EMF terms, which contribute to the growth of these large scale fields, 
are also planar averaged.  The dynamo time scale of interest in this study is 
the growth time of large scale field, not the cycle periods. 

In section \ref{result} and its subsections we present the results from the simulations.
In section \ref{planar} we discuss the presence of planar averaged large scale fields 
and distinguish the cases  of horizontal vs. vertical averaging.
Section \ref{spectra} shows the evolving magnetic and velocity power spectra mostly in the MRI growth phase.
In section \ref{terms}, we then consider planar averaged mean field equations to study and interpret the growth and saturation phase of the large scale
fields by evaluating the various terms. 
The correlation between the large scale fields and accretion stresses are discussed in section \ref{corr}.
In section \ref{discussion} we discuss the implications  for the  nonlinear saturation of the MRI dynamo, and we conclude in section 
\ref{conclusion}.

\section{Simulation setup of the MRI}
\label{simset}
We adopt the shearing box model without stratification, where the 
fluid is assumed to be isothermal, viscous, electrically conducting
and mildly compressible.
We solve the induction, Navier-Stokes and continuity equations given by,
\EQ
{\DDD\AAA\over\DDD t}=-SA_y\xxx+\UU\times\BB-\eta\mu_0\JJ \, 
\label{dAdt}
\EN
\EQ {\DDD\UU\over\DDD
t}=-\UU\cdot\nab\UU-SU_x\yyy -\frac{1}{\rho}\nab P-2({\bf \Omega}\times\UU) + \frac{\JJ\times\BB}{\rho} + \nab\cdot 2\nu\rho\mbox{\boldmath ${\sf S}$}, 
\label{dUUS} \EN 
and
\EQ {\DDD\ln\rho\over\DDD
t}=-\UU\cdot\nab\ln\rho-\nab\cdot\UU. 
\label{dlnrhoS}\EN
Here $\DDD/\DDD t\equiv\partial/\partial t+Sx\;\partial/\partial y$ includes the mean Keplerian shear flow, 
$Sx\yyy$, where $S=-3\Omega_0/2$ and the background rotational velocity is ${\bf \Omega}=\Omega_0\zzz$.
We  are solving for the deviations, $\UU$, from the Keplerian shear flow.
The magnetic field $\BB$ is related to the vector potential $\AAA$ by $\BB=\nabla\times\AAA$,
and  $\JJ=\nabla\times\BB/\mu_0$ is the current density.
In the momentum equation, the pressure $P$ and density $\rho$ satisfy $P=\rho c_{\rm s}^2$, 
where $c_{\rm s}$ is the speed of sound. The rate of strain tensor 
$\mbox{\boldmath ${\sf S}$}$ is given by
\begin{equation}
{\sf S}_{ij} = \frac{1}{2} (U_{i,j}+U_{j,i}) - \frac{1}{3} \delta_{ij} \nab\cdot\UU,
\end{equation}
where the commas denote spatial derivatives.

\begin{figure*}
\begin{center}
\includegraphics[width=0.498\textwidth, height=0.3\textheight]{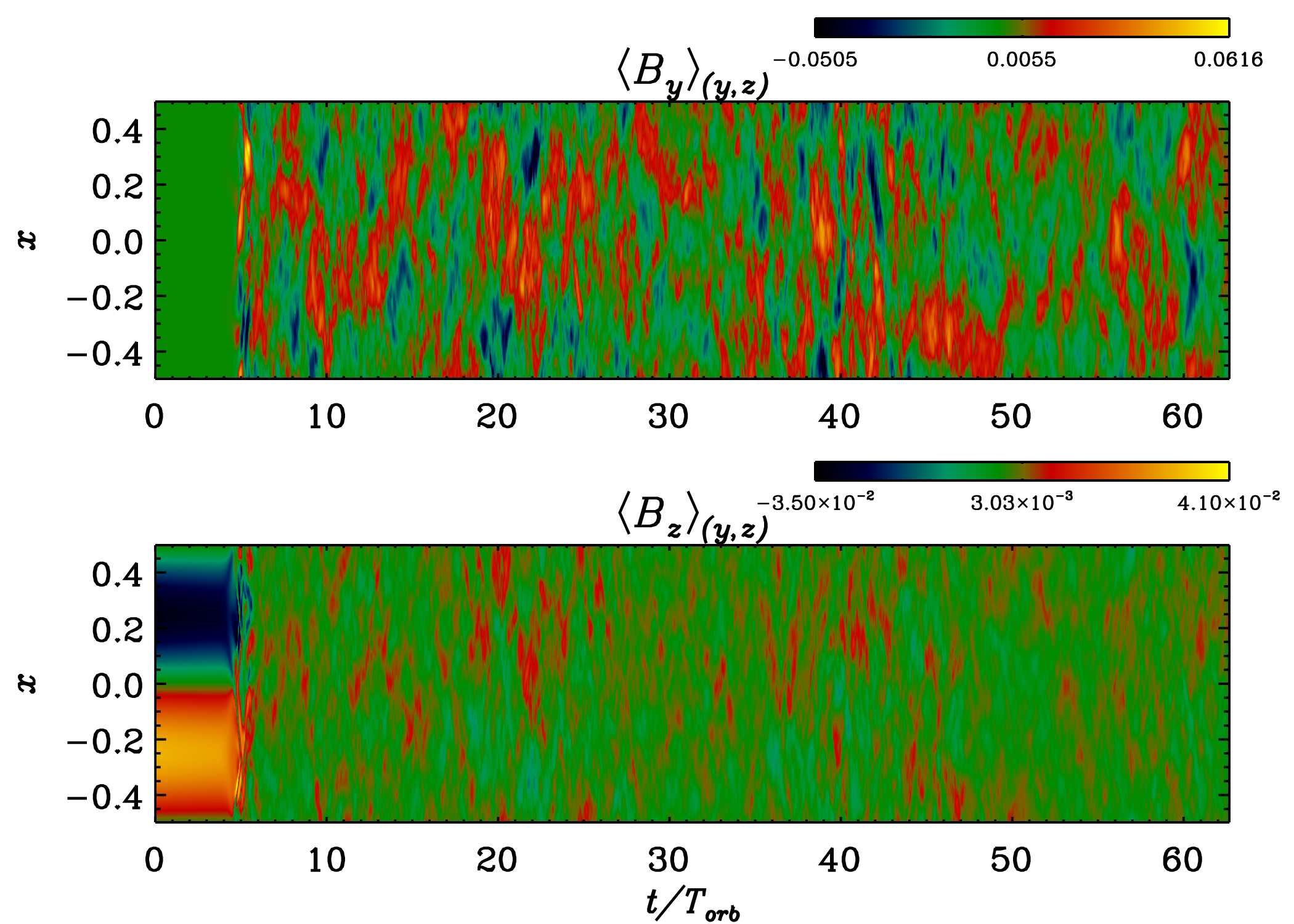}
\includegraphics[width=0.498\textwidth, height=0.3\textheight]{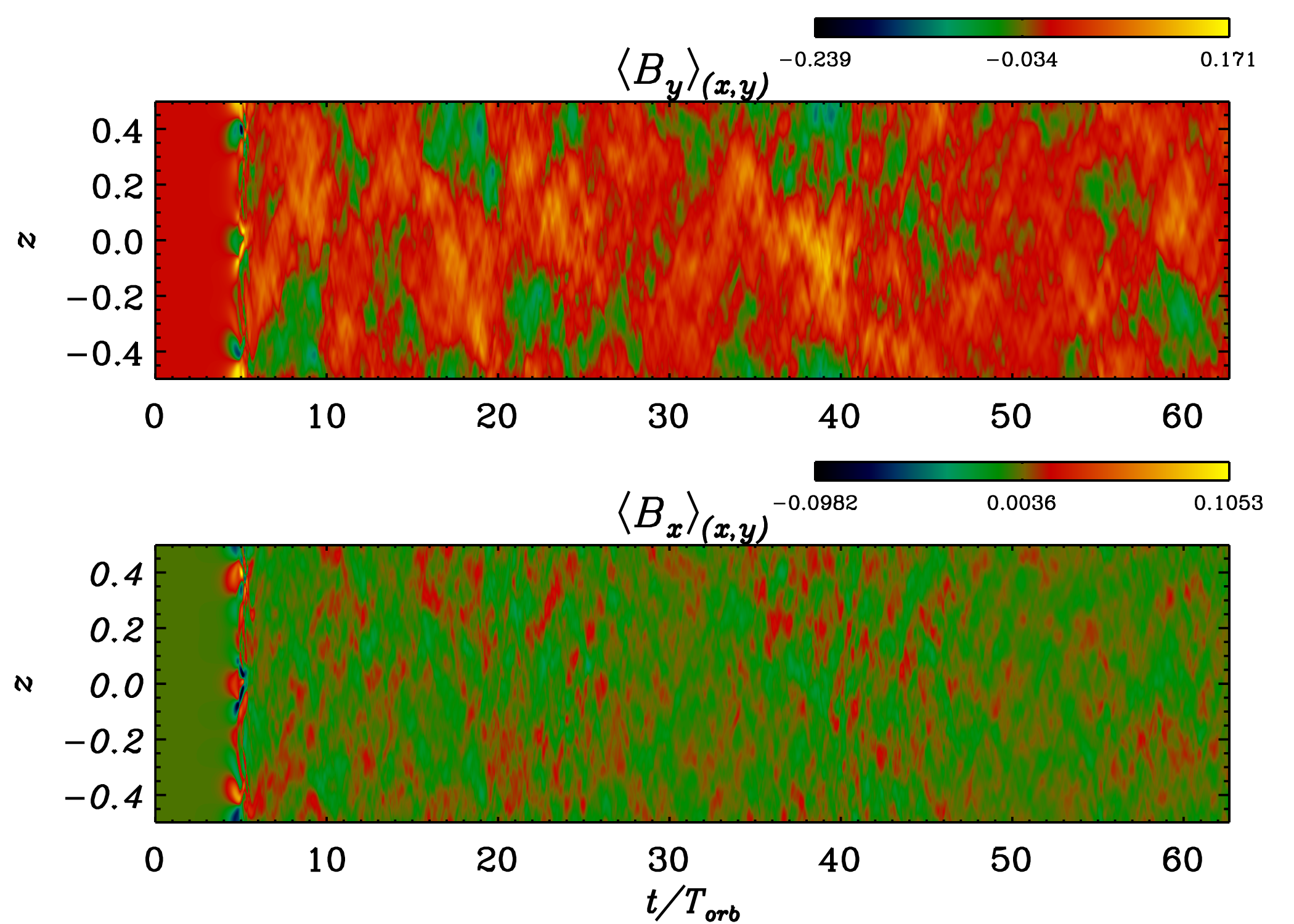}
\end{center}
\caption{Here $y$-$z$ and $x$-$y$ averaged magnetic fields are shown on the left and right respectively.  On the left, the upper panel is for 
$\overline{B}_y(x)$ and the lower panel is for $\overline{B}_z(x)$. 
On the right, the upper panel is for $\overline{B}_y(z)$ and the lower panel is for $\overline{B}_x(z)$.}
\label{horxy}
\end{figure*}

We use the
\textsc{Pencil Code}\footnote{https://github.com/pencil-code \citep{B03}},  
which is a high-order explicit finite
difference method to solve the model given in \Eqss{dAdt}{dlnrhoS}.
The model we have adopted is similar to the one used by \citet{KK11} with vertical
periodic boundary conditions. 
The boundary conditions are periodic in also $y$ and shear-periodic in $x$ \citep{WT88}. 
\Eqss{dAdt}{dlnrhoS} are solved on a $N_x\times N_y\times N_z$ Cartesian grid, 
with a size of $L_x$, $L_y$ and $L_z$ in the three Cartesian directions. We have used an 
aspect ratio of ($L_x:L_y:L_z$) = ($L:4L:L$). 
The resolutions used include ($N_x$, $N_y$, $N_z$) = ($128$, $512$, $128$) and ($256, 1024, 256$).
The code uses dimensionless quantities by measuring length in units of $L$, 
speed in units of the isothermal sound speed $c_{\rm s}$, density in units of initial value $\rho_0$ and
magnetic field in units of $(\mu_0\rho_0 c_{\rm s}^2)^{1/2}$ where $L=c_s=\rho_0=\mu_0=1$.

The initial velocity field is Gaussian random noise at the level of $10^{-3} c_{\rm s}$ 
and the initial magnetic field is given by $\BB=B_0\sin(k_x x)\zzz$, which in terms of  the vector potential  can be written as $\AAA=A_0\cos(k_x x)\yyy$, so that $\vert B_0\vert=k_xA_0$, where $k_x=2\pi/L_x$.
We choose the rotation rate, $\Omega_0=1$, and $A_0=0.005$, which results in $k_{max}/k_1=\sqrt{15/16}(\Omega_0/U_{A,0})/k_1\approx5$,
where $U_{A,0}=B_0/\sqrt{\mu_0 \rho_0}$ is the initial Alfv\'en velocity, $k_{max}$ is the wavenumber at which maximum
growth rate is expected from linear MRI analysis and $k_1=2\pi/L$.
Using the parameters above, the steady state  turbulence that develops due to the MRI
has  a characteristic  root mean square velocity
 in the steady state of
 $\Urms\sim 0.1 c_s$, ensuring that compressibility effects are small.
The ratio of thermal to initial magnetic pressure, $\beta=2\mu_0 P/{B_0}^2\approx1014$.
Also the ratio of the initial Alfv\'en velocity to the steady state root mean square, $U_{A,0}/U_{rms}\approx 0.0314/0.13=0.24$.

\begin{center}
\begin{figure*}
\includegraphics[width=0.8\textwidth]{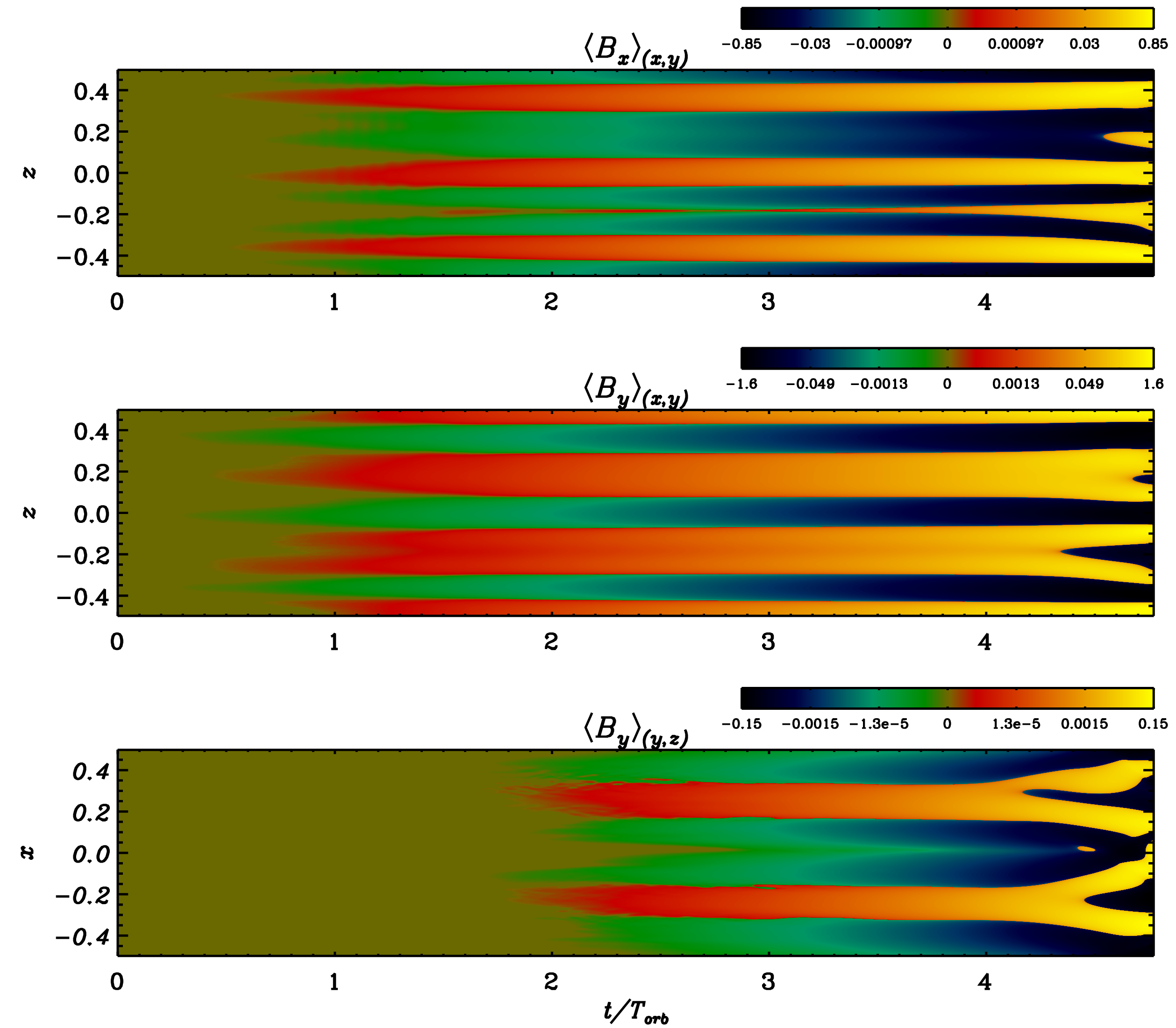}
\caption{Here $x$-$y$ and $y$-$z$ averaged magnetic fields are shown in the MRI growth phase. The top panel is for 
$\bra{\overline{B}_x}_{(x,y)}(z)$, the middle panel is for $\bra{\overline{B}_y}_{(x,y)}(z)$ and the bottom panel 
is for $\bra{\overline{B}_y}_{(y,z)}(x)$.
}
\label{linGrow}
\end{figure*}
\end{center}
We  define  the mean (or large scale) quantities as planar averages, 
so that for a given quantity $F_i$. its mean is given by
\begin{eqnarray}
\overline{{F}}_i(z,t) &=& \frac{1}{L_x L_y}\int_{-L_x/2}^{L_x/2} \int_{-L_y/2}^{L_y/2} F_i(x,y,z,t) dx dy,
\label{aver1}
\end{eqnarray}
\begin{eqnarray}
\overline{{F}}_i(x,t) &=& \frac{1}{L_y L_z}\int_{-L_y/2}^{L_y/2} \int_{-L_z/2}^{L_z/2} F_i(x,y,z,t) dy dz. 
\label{aver2}
\end{eqnarray}
We  extract two distinct set of mean quantities from the simulations by
averaging  either over the $x$-$y$ plane as in \Eq{aver1} or over the $y$-$z$ plane as in \Eq{aver2} respectively. The former results in  mean quantities being functions of $z$ 
and the latter as functions of $x$.

We also calculate the Fourier power spectra for each component of the field in each direction, given by,
\EQA
\label{aver1dspec1}
\hat{B}^2_i(k_x)&=&\bra{\vert \tilde{B}_i(k_x, y, z) \vert^2}_{(y,z)},\\
\label{aver1dspec2}
\hat{B}^2_i(k_y)&=&\bra{\vert \tilde{B}_i(x, k_y, z) \vert^2}_{(x,z)},\\
\hat{B}^2_i(k_z)&=&\bra{\vert \tilde{B}_i(x, y, k_z) \vert^2}_{(x,y)},
\label{aver1dspec3}
\ENA
where, 
\EQA
\label{1dspec1}
\tilde{B}_i(k_x, y, z) &=& \int_{-L_x/2}^{L_x/2} {B_i(x,y,z)} exp(-ik_x x) dx,\\
\label{1dspec2}
\tilde{B}_i(x, k_y, z) &=& \int_{-L_y/2}^{L_y/2}{B_i(x,y,z)} exp(-ik_y y) dy,\\
\tilde{B}_i(x, y, k_z) &=& \int_{-L_z/2}^{L_z/2}{B_i(x,y,z)} exp(-ik_z z) dz.
\label{1dspec3}
\ENA
\Eqss{1dspec1}{1dspec3} represent the one-dimensional (1D) Fourier transform for all
the 1D arrays along the different directions, $x$, $y$ and $z$ respectively.
\Eqss{aver1dspec1}{aver1dspec3} show that to obtain 1D spectra in any given direction,
we first compute the square of magnitude of the complex-valued 
1D Fourier transform, along the given direction. 
And then we average these values over the respective perpendicular planes, to obtain
the final 1D directional power spectra.
The fluid and magnetic Reynolds number are defined as $\Rey=\Urms L/\nu$ 
and $\Rm=\Urms L/\eta$ respectively, where $\nu$ and $\eta$ are the microscopic viscosity and resistivity.
Note that in the text and figures, we use overline, $\bar{ }$ and brackets, $\bra{ }$ alternatively
to indicate spatial averaging (mostly planar averaging).
\begin{figure}
\includegraphics[width=0.48\textwidth, height=0.3\textheight]{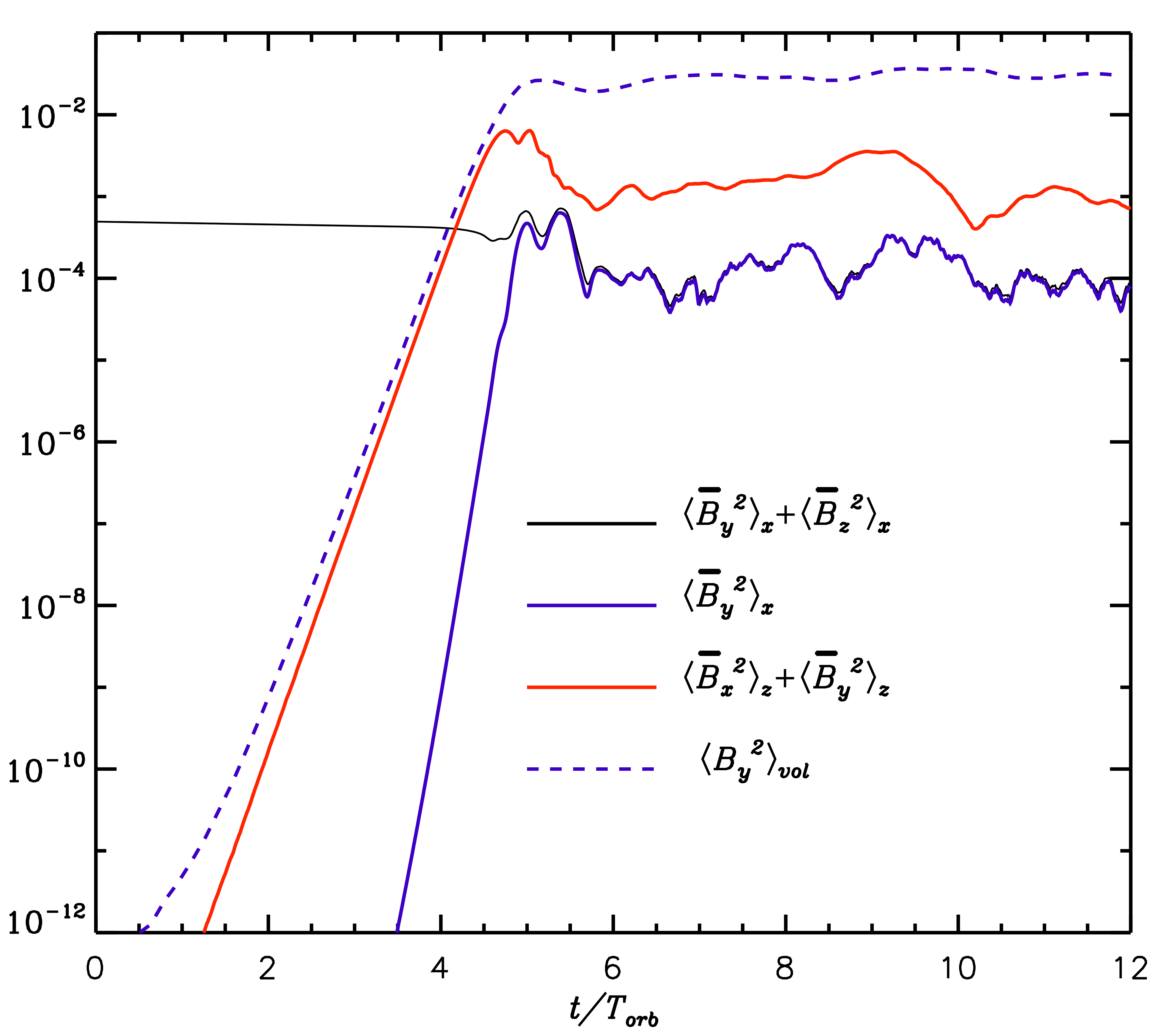}
\caption{We indicate here the evolution of the planar averaged mean field energy
densities for Run~A.
The solid black line shows the sum of squares of the vertically averaged mean fields  :$\bra{\overline{B}_y^2} + \bra{\overline{B}_z^2}$. 
The solid red line shows the sum of squares of the horizontally averaged mean fields :$\bra{\overline{B}_y^2} + \bra{\overline{B}_x^2}$. 
The blue line shows the evolution of only the the vertically averaged azimuthal mean field and the dashed
blue indicates the energy in the azimuthal component of total field.}
\label{meanevol}
\end{figure}
\section{Results}
\label{result}
The results here are from direct numerical simulation (DNS) runs with  resolutions of $256\times1024\times256$ and $128\times512\times128$, 
denoted by Runs~A and B, respectively. Both  runs have the same $\Rm=1250$
and Prandtl number $\Pm=\Rm/\Rey=4$. In  \Fig{timeplot}, we plot the evolution of $U_{rms}$ and $\Brms$ from the two 
runs which shows that both incur a similar strength.  We later show that  the stresses for the two runs are also similar.  
We therefore consider the results to be suitably converged and take Run~A, which has the higher resolution, to be our fiducial run.
Note that the growing $\Brms$ indicates MRI growth phase upto $t/T_{orb}\sim5$ and then $\Brms$ settles into steady state 
indicating the nonlinear saturation regime.
We discuss the results from Run~A in more detail in what follows.

\subsection{Planar averaged large scale or mean fields}
\label{planar}
We discuss results for the two  different planar averages, $x$-$y$ (horizontal)  and $y$-$z$ averaging (vertical) and 
determine the respective mean magnetic fields  $\meanBB(z)$ or $\bra{\BB}_{x,y}$ and  $\meanBB(x)$ or $\bra{\BB}_{y,z}$.

In the right two panels of \Fig{horxy}, the strengths of the  $x$-$y$ averaged fields  are
indicated by the color scaling and seen to evolve with time along the  abscissa
and vary in   $z$ along the ordinate. A strong large scale field  $\overline{B}_y(z)$, is seen after $t/T_{orb}\sim 5$ in the non-linear turbulent regime. 
The large scale field $|\overline{B}_x(z)|$ is smaller than  $|\overline{B}_y(z)|$ by a factor of 2 and somewhat less coherent.
\begin{center}
\begin{figure*}
\includegraphics[width=\textwidth]{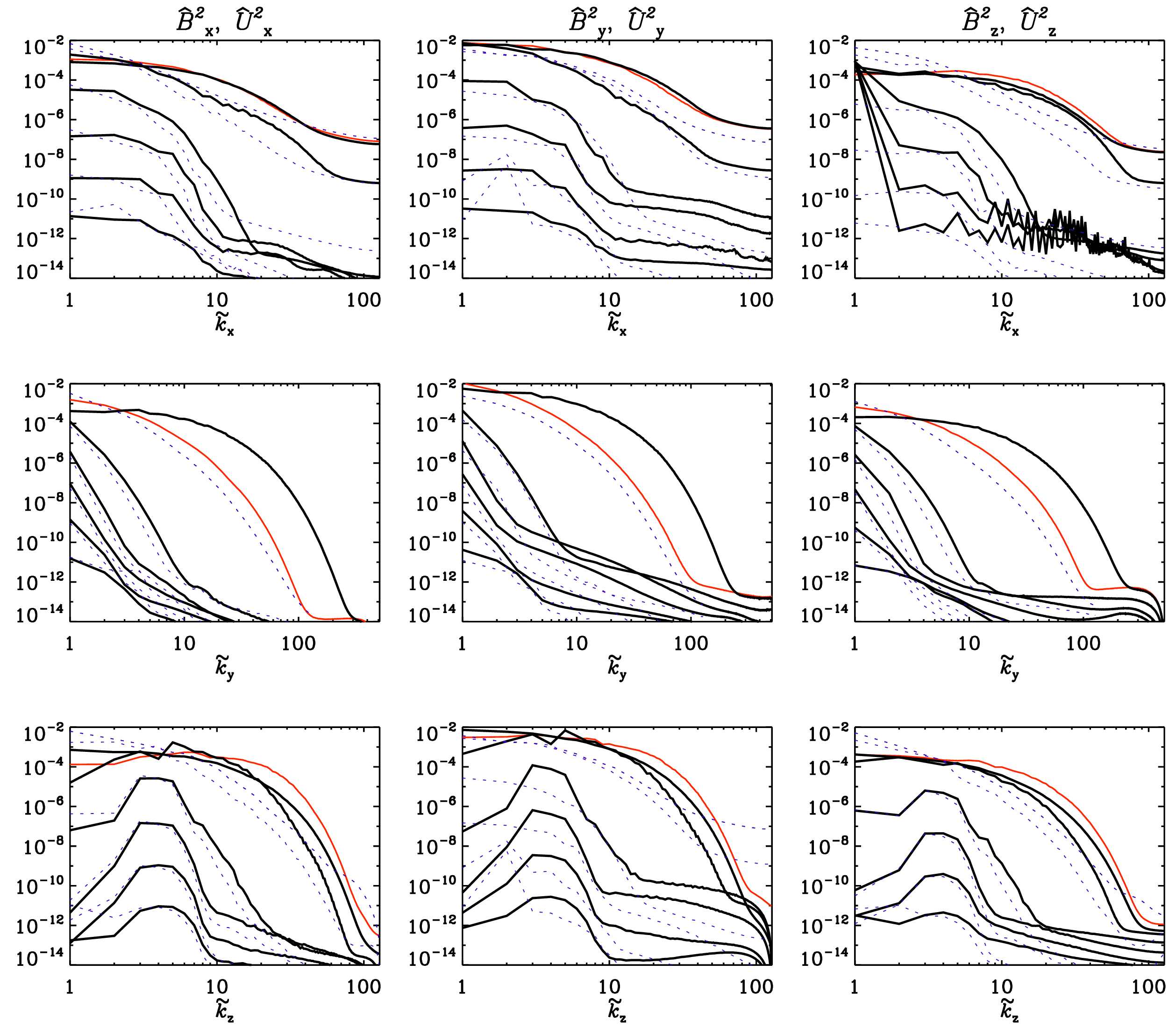}
\caption{Evolution of 1D Fourier spectra for each component of the magnetic and velocity fields are shown in solid black and dashed
blue respectively. Each of the three components of the fields,  $\{\hat{B}_x, \hat{u}_x\}$, $\{\hat{B}_y, \hat{u}_y\}$, $\{\hat{B}_z, \hat{u}_z\}$ 
are along the three columns left to right respectively and the rows depict the spectra along the three wavenumbers, $\tilde{k}_x,\tilde{k}_y,\tilde{k}_z$,
from top to bottom respectively.
The spectra in each panel are shown at equidistant intervals, $t/T_{orb}=1.58, 2.38, 3.18, 3.98, 4.78, 5.58$. The field saturates by $t/T_{orb}\sim5$.
And the final curve in saturated regime for $\hat{\BB}$ is shown in red.
Particularly in the last row, note that the spectra are initially peaked at $3\le \tilde{k}_z\le 4 $, which is nearly $u_{A,0}/\Omega_0$, in the MRI growth phase.
}
\label{spec}
\end{figure*}
\end{center}
Previous studies of MRI shearing box simulations  have employed horizontal averaging to compute mean fields.
However, recent global DNS of MRI for a cylinder,  using periodic boundary conditions in $z$ and perfectly conducting walls for the innermost
and outermost radii, have employed vertical averaging \citep{ebrahimiprl,EB16} 
over $z$ and  $\phi$ (or azimuthal) directions to compute mean fields as functions of radius.
We therefore want to assess how results from shearing boxes compare  with results 
from the global cylinder simulations when the same averaging is used.
 For our shearing box simulations, vertical ($y$-$z$) averaging does indeed produce a strong large scale field $\overline{B}_y(x)$, as seen in the left upper panel of \Fig{horxy}. 
The mean field $\overline{B}_z(x)$ (seen in the left lower panel in \Fig{horxy}) in the growth phase ($t/T_{orb}\lesssim 5$) reflects the initial condition of $B_z=B_0\sin(k_x x)$,
although the vertical field becomes more turbulent upon nonlinear saturation.
The field $\overline{B}_y(x)$ is stronger than $\overline{B}_z$ by a factor $\sim 3$. This is also the case in the global DNS of EB16.
We therefore find that  the shearing box results are consistent with the results from the global DNS.

\begin{figure}
\includegraphics[width=0.495\textwidth, height=0.25\textheight]{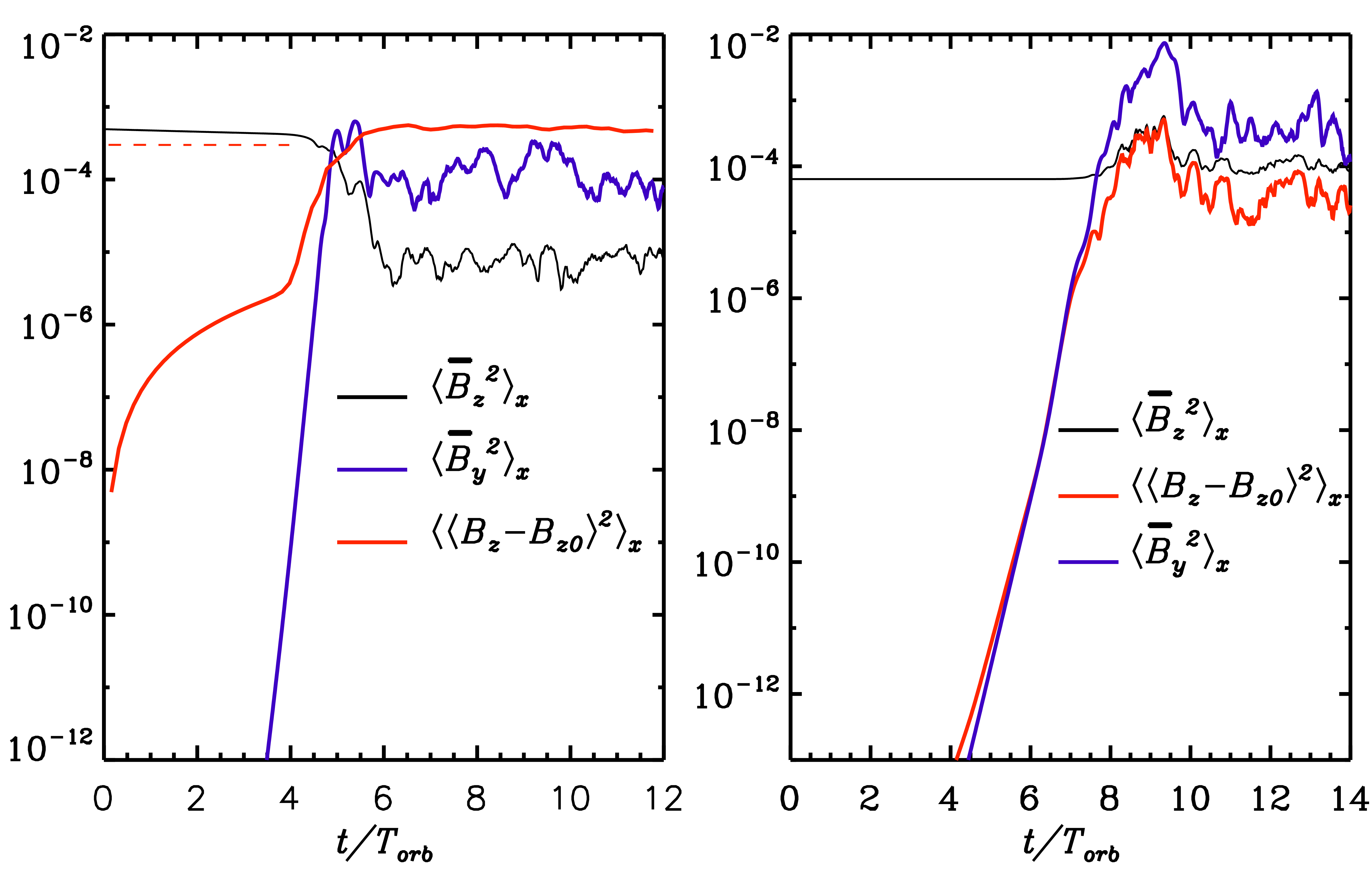}
\caption{We show here the evolution of the mean square of $y$-$z$ averaged mean fields.
In the left panel, the solid black line shows the total $\overline{B}_z^2$ and blue line shows $\overline{B}_y^2$
for Run~A. The red dashed line indicates the slope expected for resistive decay of the field at $\tilde{k}_x=1$
(also the initial field). The solid red line is for the quantity, $\bra{(\overline{B}_z-\overline{B}_{z0})^2}$
In the right panel, $\overline{B}_z^2$ and $\overline{B}_y^2$ is shown in red
and blue lines for a run where the initial field is an imposed constant and uniform mean field $B_0=0.008$. Additionally,
we show the curve for $\bra{B_z-B_0}^2$, indicating growth of mean field independent of the background field.}
\label{bzcomp}
\end{figure}
Prominently seen in the  $x$-$y$ averaged plots in the right panel of \Fig{horxy}
around $t/T_{orb}\sim 5$, are short loops, just before $\Brms$ turns to saturate. 
These features are  indicative of the modal quasilinearity in the MRI growth phase, 
but are not easily  seen  in the growth phase due to the linear scaling of the contour colours. 
To reveal these  features more conspicuously, 
we show in \Fig{linGrow}, the averaged magnetic field evolution restricted to the MRI growth phase, where the contour
colours scale logarithmically, highlighting the exponential growth of the field
components orthogonal to the initial field. 
The figure reveals  persistent 
large scale field growth right from the start of the MRI that continues 
into the nonlinear  saturation regime.
By counting the number of yellow and blue bands (which form the crests and troughs
of the fastest growing mode at $k_{max}$) in any of the upper two panels of \Fig{linGrow}, 
we can determine that  $k_{max}/k_1\sim4$, which is slightly less
than the estimate  $(\Omega/V_A)/k_1$. 
Similarly, in the $y$-$z$ averaged large scale field, $\overline{B}_y(x)$, we find a large scale mode with extended
coherence in the $x$ direction. Thus, these plots reveal the presence of growing large scale fields
and are consistent with quasilinear nature of large scale field growth in EB16.

To elucidate the growth of the planar averaged fields in the MRI growth phase,
we show  the sum of the mean square of planar averaged fields in \Fig{meanevol}.
In the saturation regime, the energy in horizontally averaged fields is seen
to be higher than that in vertically averaged fields. However, all of these curves
show that components of the large scale fields already grow the  early MRI growth phase, and are sustained in the nonlinear saturation regime.
As we shall  see later,  the  $x$-$y$ averaged mean field $\overline{B}_x(z)$ arises  from the  electromotive force (EMF) in the 
mean field induction equation while the  $\Omega$-effect 
amplifies  $\overline{B}_y(z)$. In the $y$-$z$ averaged equations, the 
$\Omega$-effect operating on the mean field is absent because  $\bra{B_x}_{(y,z)}(x)$ is zero (as a result of  the field being divergence-free).
Thus both $\overline{B}_y(x)$ and $\overline{B}_z(x)$ arise due to their respective EMF terms
in the induction equation.
This compares well with EB16, who calculated the EMFs responsible for the dynamo
production of vertically averaged, radially varying mean fields in a quasilinear calculation.
In Sec. \ref{terms} we explicitly evaluate
terms in the mean field equations  and discuss their
contributions further. To better study the growth of fields in early phase, we turn to spectral methods.

\subsection{Evolution of magnetic power spectra}
\label{spectra}

In previous work, the MRI has been discussed mainly by studying the distribution of the magnetic energy across vertical wave numbers.  
But if turbulent diffusion ensues in the non-linear regime, then the 
sustenance of a large scale field 
requires a mechanism of exponential growth to compete with the turbulent exponential decay.
This in turn requires field amplification in more than one mutually perpendicular direction.

\Fig{spec} shows the evolving power spectra for all components $\hat{B}_x^2$, $\hat{B}_y^2$ or $\hat{B}_z^2$
(also $\hat{U}_x^2$, $\hat{U}_y^2$ or $\hat{U}_z^2$) as given by \Eqss{aver1dspec1}{aver1dspec3}, 
along $\tilde{k}_x=k_x/k_1$, $\tilde{k}_y=k_y/k_1$ and $\tilde{k}_z=k_z/k_1$ during 
the linear amplification stage of MRI and up to saturation. 
The power spectra for all components of magnetic and velocity fields as a function of  $k_z$ (bottom row)
exhibit a peak corresponding to $\tilde{k}_z=k_{max}$,  resulting from the fastest growing mode of the MRI. 
(The fastest growing mode of the MRI is usually  discussed with respect to  vertical wave numbers.)  
Also noteworthy is the predominance of low wavenumber modes in $\tilde{k}_x$ and $\tilde{k}_y$. This includes  not just the 
analytically expected  $\tilde{k}_x=\tilde{k}_y=0$  axisymmetric modes
but also $\tilde{k}_x,\tilde{k}_y \in [1$,$n]$, where $n$ is a small integer 
$ > 1$ which differs for $\tilde{k}_x$ and $\tilde{k}_y$.
These low wave number modes grow for all field components.
The growing modes in $\tilde{k}_x,\tilde{k}_y$ are confined to 
first few wave numbers (particularly narrow in the case along $\tilde{k}_y$).

From the bottom panel of \Fig{linGrow}, note that the number of bands in $\bra{B_y}_{y,z}$ along $x$, 
corresponds to the mode $\tilde{k}_x=2$ in $\hat{B}_y^2(\tilde{k}_x)$ (seen in the top middle panel in \Fig{spec})
as this is the dominant mode; it is  energetically ($\tilde{k}_x \hat{B}_y^2$) larger than the others.
A similar correspondence manifests for the mean fields, $\bra{B_x}_{x,y}$ along $z$ and $\bra{B_y}_{x,y}$ along $z$
with $\hat{B}_x(\tilde{k}_z)$ and $\hat{B}_y(\tilde{k}_z)$ respectively, with the dominant mode at $k_{max}$.
Thus the modal structure of the planar averaged mean fields is determined by the
dominant modes in the power spectrum.
From both \Fig{linGrow} and  \Fig{spec}, we  find that (i)  large scale
field components grow right from the early MRI growth phase
 and (ii) the fields reside in a narrow set of modes which are not yet turbulent.
Thus, the large scale dynamo action in the MRI growth phase does not  require turbulence, which is  consistent with the quasilinear analysis of EB16.
In the MRI growth regime, there is not yet turbulent diffusion of the initial ${\overline B}_z(x)$ to overcome, but
the low wavenumber modes do grow in $\tilde{k}_x$ for all components (including $B_z$).
In the top right panel of  Fig \ref{spec}, the evolving power spectra for $\hat{B}_z(\tilde{k}_x)$  
indicate that the amplitude of the $\tilde{k}_x=2$  mode increases with time.
Thus the instability driven low wavenumber modes feed back directly into the initial vertical field.

\subsubsection{Analysis of the MRI growth regime up to saturation}

We now analyse how this feedback affects the vertical mean field by examining the evolution 
of the $y$-$z$ averaged mean field $\overline{B}_z(x)$. The left panel of \Fig{bzcomp}, is our fiducial run 
which starts with a mean vertical field of zero net flux.
The panel shows  that in the MRI growth regime,  the mean vertical
field (black curve)  actually  decays  faster than resistive decay rate
for $k_x=1$ (where the resistive decay rate $\sim \eta \tilde{k}_x^2$), 
with the latter  shown in dashed red. The  faster than resistive decay rate arises because 
the growing low wavenumber MRI modes $\tilde{k}_x\in [1$,$5]$ allow for more reversals in the mean field
so that the effective $\tilde{k}_x>1$ in a more accurate estimate of the resistive decay rate.
Since the vertical field evolution in the growth regime is dominated by the initial $\tilde{k}_x=1$ mode,
the evolution of the mean vertical field energy density 
shown by the black curve reflects mainly the dynamics of this  mode.
However, as mentioned earlier, from the spectra associated with low wavernumber modes in $\tilde{k}_x$, we expect a 
growing mean vertical field which is hidden as these modes grow from small  initial amplitudes.

To further explain how such  growing  modes of the mean field  $\overline{B}_z(x)$ can be hidden, 
we consider a complementary simulation in which  we start with a 
constant and uniform background vertical field
(and thus a net vertical flux), where initially there is no power in any of the higher modal scales. 
The right panel of \Fig{bzcomp}, shows two curves, one for the total
mean vertical field (in black) and one for which the constant imposed field is subtracted from the total and the result is vertically averaged
to obtain a growing mean field (in red). Thus, this panel shows that  the growth of MRI unstable low wavenumber modes in $\tilde{k}_x$, 
contributes to the growth of a vertical mean field, thus adding to the background field. 

The zero-net-flux left panel of \Fig{bzcomp} is different from the  net-flux case of the right panel
because we start with so much power in  a mode that could otherwise grow, that the 
MRI modes cannot grow to an amplitude that exceeds the initial value before the onset of
saturation phase .
Thus, we see that 
the dominant initial mode at $\tilde{k}_x=1$ in the mean vertical field 
actually decays from its initial value.
However subsequently it settles into a steady state, which indicates
that these mean field  MRI modes are in fact growing behind the scenes, 
balancing turbulent diffusion.
To show this more explicitly, we plot the quantity, $\bra{(\overline{B}_z(x)-\overline{B}_{z0}(x))^2}$, where $\overline{B}_{z0}$
is the initial vertical mean field,  and  $\bra{B_z-B_{z0}}=\overline{B}_z-\overline{B}_{z0}$. 
The evolution of this quantity is shown in the red curve in the left panel of \Fig{bzcomp}.
The red curve initially rises rapidly, indicating the presence of the small amplitude growing MRI modes which average to  a finite mean.  
Upon saturation this curve rises again and flattens reflecting the turbulent diffusion of the dominant mode and eventual 
steady state due to the competition between growing modes and turbulent diffusion.
Note that towards the end of the growth phase the red curve asymptotically matches with the amplitude of the total vertical 
mean field in black curve,  supporting the idea that the growing 
MRI unstable modes feed back into the vertical mean field and 
contribute to the steady state mean field sustenance
against  turbulent diffusion. 

The two panels  of  \Fig{bzcomp} 
may in fact be mutually consistent in that the left panel starts with a large initial mean field in the modes
that would otherwise  show growth, rather than decay, if initiated at a much lower value.
In the right panel, these modes visibly grow exponentially up to their saturated value.
In the simulations to date,  it is difficult to reduce the strength of the initial field to much lower
values, as then the $k_{max}$ moves to large values where damping due to explicit viscosity and resistivity becomes significant.
The saturated regime of the two panels is qualitatively similar, and shows that this 
direct feedback from the instability-grown low wavenumber modes (which survive averaging) to the initial vertical field
potentially sustains the field against turbulent decay in the saturation regime.

\begin{center}
\begin{figure*}
\includegraphics[width=0.8\textwidth, height=0.4\textheight]{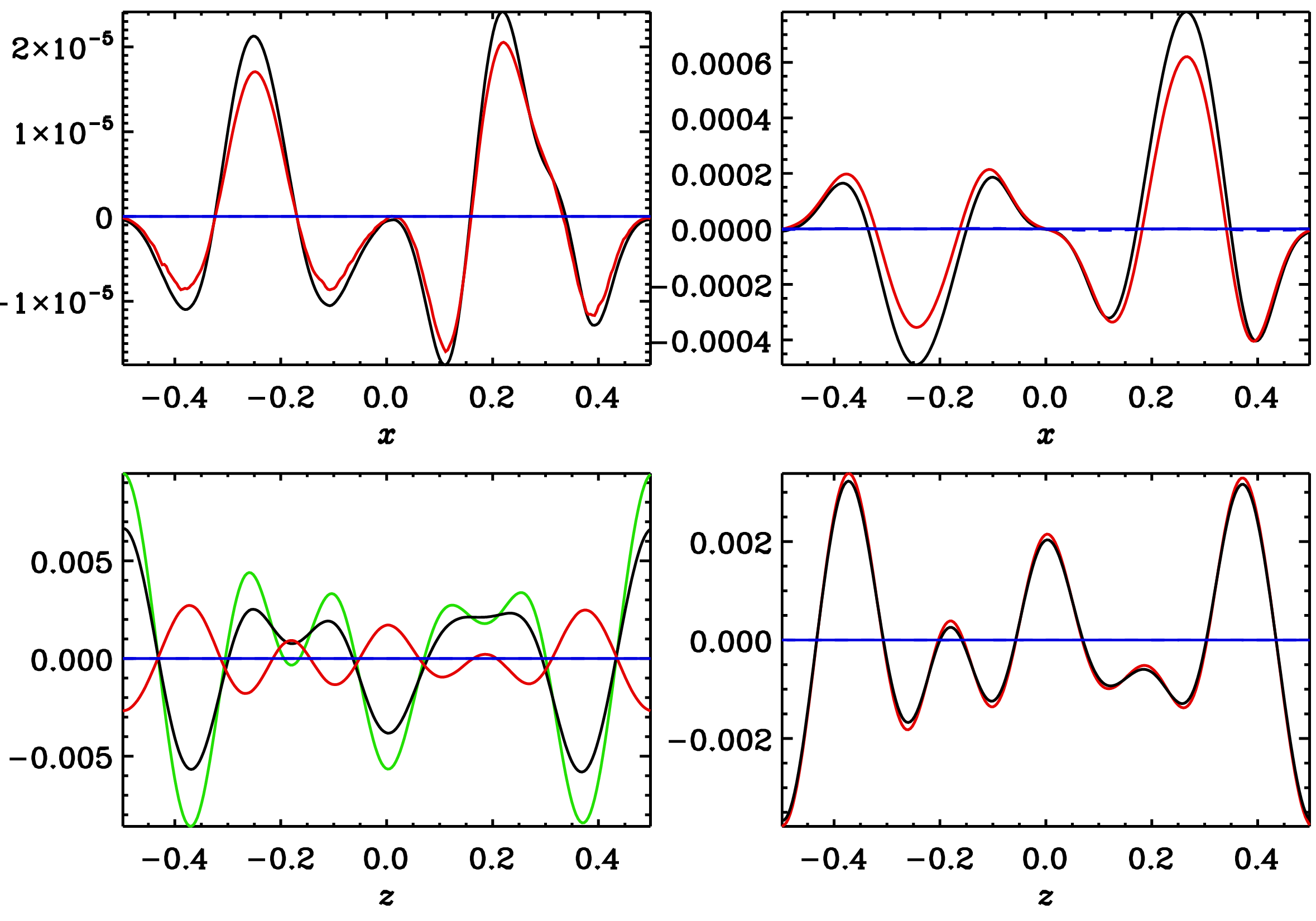}
\caption{The top two panels show the terms from $y$-$z$ averaged mean field equation for $\overline{B}_y(x)$ and $\overline{B}_z(x)$
on left and right respectively. The bottom two panels show the terms from $x$-$y$ averaged mean field equation 
for $\overline{B}_y(z)$ and $\overline{B}_x(z)$ on left and right respectively. These are plotted at $t/T_{orb}\sim4$, in the
MRI growth phase. The solid black curve is for the time derivative of the mean field, the red curve is for the corresponding EMF
term, the green curve is for the term $S\overline{B}_x$, and the blue solid and dashed lines are respectively for the advection and stretching terms
involving mean fields.
}
\label{mrimft1}
\end{figure*}
\end{center}
\begin{center}
\begin{figure*}
\includegraphics[width=0.8\textwidth, height=0.4\textheight]{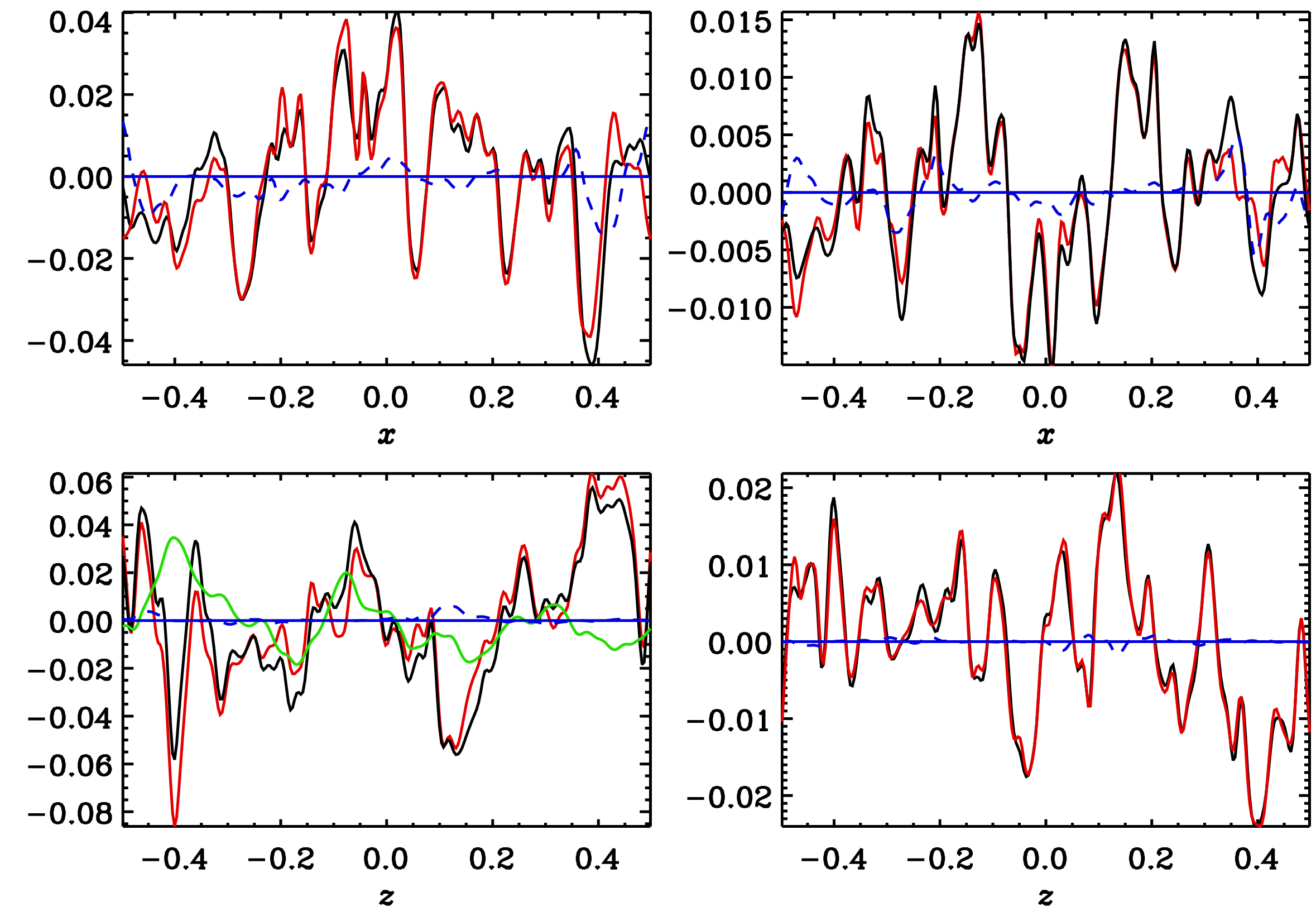}
\caption{The top two panels show the terms from $y$-$z$ averaged mean field equation for $\overline{B}_y(x)$ and $\overline{B}_z(x)$
on left and right respectively. The bottom two panels show the terms from $x$-$y$ averaged mean field equation 
for $\overline{B}_y(z)$ and $\overline{B}_x(z)$ on left and right respectively. These are plotted at $t/T_{orb}\sim19$, in the
MRI saturation phase. The solid black curve is for the time derivative of the mean field, the red curve is for the corresponding EMF
term, the green curve is for the term $S\overline{B}_x$, and the blue solid and dashed lines are respectively for the advection and stretching terms
involving mean fields.
}
\label{mrimft2}
\end{figure*}
\end{center}
The growth and sustenance of large scale toroidal magnetic energy seen in 
\Fig{bzcomp}~(a) is similar and consistent with the results of zero-net flux 3D cylindrical simulations in EB16 (see Fig.~2).  
Also, a similar effect of MRI feedback on the initial vertical field was found in the cylindrical simulations of nonzero-net flux \citep{ebrahimi2009}. 
It was there shown that the initial vertical field is amplified through a single mode azimuthal EMF, resulting in the saturation of the mode. 
These similar features between global cylindrical and the shearing box simulations presented here, 
confirm the robust nature of large scale dynamo growth during MRI early phase.

Close to  saturation, high wavenumber modes arise and the power spectra (both magnetic and velocity) broaden.
Also subsequently a peak at $\tilde{k}_z=1$ rises for $\overline{B}_z$.
This is best  interpreted as a shift in the fastest growing mode towards lower
wavenumbers, as the field has grown leading to a larger Alfv\'en velocity
and smaller $k_{max}\sim \Omega/V_A$.

\subsubsection{Saturated regime}
On saturation,  the broad power spectra suggest that  turbulent diffusion would ensue
and the growing low wavenumber modes which survive $y$-$z$ averaging  potentially compete with turbulent diffusion to support
sustenance of the mean vertical field, which is important for the sustenance of MRI turbulence.
In saturation, the power spectra for all components for all the different wave numbers, 
remain broad, with the curves sloping down from low to high wavenumbers monotonically.

Overall, the spectral evolution is consistent with the interpretation that the MRI first grows low wavenumber modes
which sustain the original field responsible for driving these modes, 
establishing a self-sustaining, instability-driven large scale dynamo.
Note that here in the MRI dynamo, the large scale fields arise first, 
and are only later followed by turbulence  (possibly due to the tangling of the large scale fields).
This is conceptually different from the case of large  $\Rm$ 
helical dynamos forced with small scale turbulence in a box \citep{BSB16}, where the magnetic energy on all scales grows at the same rate,
with most of power peaked close to the small resistive scales first.
We study these results using mean field theory in our next section.

\section{Evaluating terms in mean field equations}
\label{terms}
We present here the mean field equations obtained from $x$-$y$ and $y$-$z$ averaging.
The fields can be split into a mean component and a fluctuating component, given by,
${\UU}=\meanUU+\uu$ and ${\BB}=\meanBB+\bb$.
The mean field equations in $x$-$y$ averaging are given by,
\EQA
\label{xyeq}
\partial \overline{B}_x \over \partial t &=& - \partial_z \emf_y + \overline{B}_z \partial_z \overline{U}_x - \overline{U}_z \partial_z \overline{B}_x \\ 
\partial \overline{B}_y \over \partial t &=& S\overline{B}_x + \partial_z \emf_x + \overline{B}_z \partial_z \overline{U}_y - \overline{U}_z \partial_z \overline{B}_y.
\ENA
where $\emf_y=\bra{u_z b_x- u_x b_z}$ and $\emf_x=\bra{u_y b_z- u_z b_y}$ are different components of the EMF ${\bf \emf}$.
The mean field equations in $y$-$z$ averaging are given by,
\EQA
\partial \overline{B}_y \over \partial t &=& - \partial_x \emf_z + \overline{B}_x \partial_x \overline{U}_y - \overline{U}_x \partial_x \overline{B}_y \\
\partial \overline{B}_z \over \partial t &=& \partial_x \emf_y + \overline{B}_x \partial_x \overline{U}_z - \overline{U}_x \partial_x \overline{B}_z.
\label{yzeq}
\ENA

In \Figs{mrimft1}{mrimft2}, 
we show the individual terms in \Eqss{xyeq}{yzeq}, for both $y$-$z$ and $x$-$y$ averaging (first and second rows respectively).
\Fig{mrimft1} corresponds to the MRI growth regime, evaluated at $t/T_{orb}=4$ and \Fig{mrimft2} 
corresponds to the nonlinear saturation regime, evaluated at $t/T_{orb}=19$.
The time derivative of field $\partial \meanBB/ \partial t$ 
is plotted in black, the corresponding EMF term is in red, the
shear term ($S\overline{B}_x$) is in green,
the stretching term ($\meanBB\cdot\nabla\meanUU$) is in solid blue and the advection term ($\meanUU\cdot\nabla\meanBB$) is in
dashed blue line, where the last two are negligible.
These terms are varying along $x$ or $z$, for $y$-$z$ and $x$-$y$ averaging respectively.

In the linear/quasilinear phase as well in the nonlinear regime, the mean fields,
$\bra{B_x}_{(x,y)}$ (bottom right panel), $\bra{B_y}_{(y,z)}$ (top left panel) 
and $\bra{B_z}_{(y,z)}$ (top right panel), seem to fully arise from the EMF.
However $\bra{B_y}_{(x,y)}$ results from a combination of the shear term and the EMF term. In the MRI growth regime,
the shear terms seem to be larger, but in the nonlinear regime the EMF is seen to dominate at the given instance in time. 
There is conspicuous growth of large scale fields with large scale modes
in the MRI growth phase, which can also be viewed
from the mean field equations, indicative of  large scale dynamo action. 
The results of $y$-$z$ averaging are consistent with  the results obtained in global MRI simulations of EB16, showing
that $\overline{B}_y(x)$ arises due to EMF alone as opposed to the case of $\overline{B}_y(z)$.
In the MRI growth regime, the EMF results from  fluctuations but 
which have not yet achieved a turbulent state as the contributing modes reside only on a narrow range as seen in \Fig{spec}.
The EMF here is defined
according to the standard mean field induction equation (as in \citet{Moffatt}) and is different from the definitions in
\citet{LesurOgil2008} and \citet{herault2011}, thus a comparison is not possible. 

\subsection{Growth rates of the mean fields}
We have estimated the growth rate of the mean fields in both types of averaging. 
The growth rate is estimated as $\lambda=\bra{\partial \ln\overline{\BB}^2/\partial t}/2$, where the 
the brackets denote averaging over the 1D domain in which the planar averaged field varies. 
In the case of $x$-$y$ averaging, the normalised growth rate $\lambda T_{orb}$ for both $\overline{B}_x(z)$ and $\overline{B}_y(z)$ is $\sim 2\pi*0.53$.
The linear MRI dispersion relation for vertical wavenumbers is given by 
\EQ
(\omega/\Omega_0)^2 = p^2 + (2-q) - \sqrt{(2-q)^2 + 4p^2}
\label{disp}
\EN
where $\omega$ is the growth rate of any vertical mode, $p=\tilde{k}_z U_{A,0}/\Omega_0$ and $q=-S/\Omega_0=1.5$.
From \Eq{disp}, for $\tilde{k}_z=\Omega/U_{A,0}$, we have $p=1$ and $\omega=0.56\Omega_0$ or $\omega T_{orb}=(2\pi)\times0.56$. 
In the simulation, we obtain a similar growth rate for the vertical peak mode, $\tilde{k}_z=4$, to be $\sim (2\pi)\times0.5$. Thus the growth rates
of the $x$-$y$ averaged mean fields are comparable to growth rates of the fastest growing MRI unstable modes.
Interestingly, in the case of $y$-$z$ averaged field, $\overline{B}_y(x)$, growth rate is $\sim (2\pi)\times0.91$, 
which is larger than that of the fastest growing vertical modes. 
Thus the planar averaged mean fields in MRI growth phase are found to be growing on time scale of the instability itself. 

\subsection{Correlation between stresses and mean magnetic field}
\label{corr}
\begin{figure}
\includegraphics[width=0.495\textwidth, height=0.36\textheight]{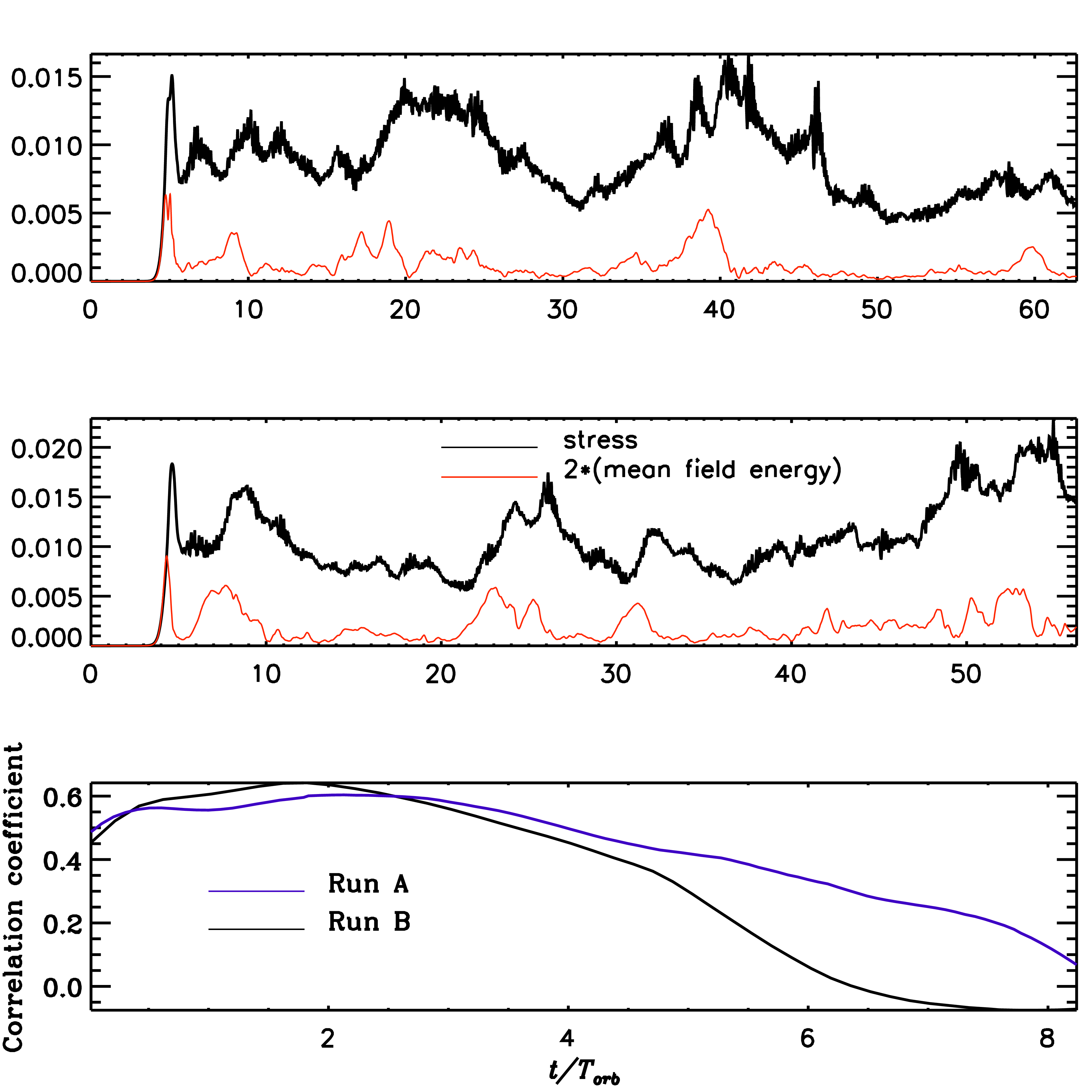}
\caption{We show here the evolution of the stresses and the large scale fields and the correlation 
between them. Top panel is for Run~A and the middle panel is for Run~B. In the bottom panel, we show
the correlation coefficient as a function of time.}
\label{stress}
\end{figure}

The most common application of the MRI is the transport of angular momentum in accretion disks via the associated   Maxwell  and Reynolds stresses \citep{BH91}.
More recently, it has been seen in  simulations of the MRI that the sustenance of
the turbulence and stresses correlates with sustenance of the large scale field \citep{LOg08,DSP10}.
It is therefore germane to assess whether there is a correlation between the dynamo generated 
magnetic mean field and the stresses in our present work. In the upper two panels of \Fig{stress}, 
the energy of the $x$-$y$ averaged mean field, given by $E(t)={\overline{B}_x}^2 + {\overline{B}_y}^2$, 
in red can be seen to correlate with the total stress given by, $S(t)=\bra{u_x u_y}-\bra{B_x B_y}$ in black.
In the bottom panel of \Fig{stress}, we show the estimate of of the linear Pearson correlation coefficient,
given by $R=$cov$(E(t),S(t-\tau))/(\sigma_E\sigma_S)$, where cov$(E(t),S(t-\tau))$ is the covariance and $\sigma_E$, $\sigma_S$
are the standard deviations for $E(t)$ and $S(t-\tau)$ respectively. The correlation coefficient 
is seen to peak at a time lag of $\tau=t/T_{orb}\sim2.1$ with a value of $R=0.60$ for Run~A and for Run~B, 
$R$ is seen to peak at a time lag $\tau=t/T_{orb}\sim1.8$, with a value of $R=0.63$.
This further highlights the importance of studying how large scale fields evolve and saturate,
which may also be essential to understanding how the stresses  of MRI turbulence saturate.

\section{Implications for the Nonlinear Saturation of the MRI}
\label{discussion}

Our focus in the present paper is  on the onset of large scale field growth, 
but  studying the spectral evolution reveals a plausible   phenomenology of  saturation to 
be further understood.
\Fig{spec} shows  that close to the onset of nonlinear saturation, small scale fields
appear which broaden the power spectra for all components along all wave numbers.
In particular for the bottom row of panels, 
the field is  initially peaked at the initial $k_{max}$, 
but at later times, a peak at $\tilde{k}_z=1$ arises. This seems to indicate the expected shift
of the fastest growing mode as the  field strength increases. 
Thus while other modes may have stabilized, the field continues to grow at the largest scales, 
competing with turbulent diffusion. 

Such an evolution toward saturation  provides a clue as to why the 
Maxwell to Reynolds stress ratio exhibits the same dependence on  the shear parameter, 
$q$ (equal to $-S/\Omega_0$ in our model)
during both the MRI linear and saturation phases \citep{PCP06}:
The modes at largest scales start as linear modes and continue as such longer than the small scale modes
even when turbulent dissipation ensues.

Note that both velocity and magnetic fields grow and saturate together with similar
power spectra throughout the simulations. This we expect from linear theory, where both velocity and magnetic fields
grow with the same eigenfunction.
In the saturated regime, both velocity and magnetic fields  peak at the largest scales.  
We can question whether such a dynamo can be called a large scale dynamo, given that the effective forcing scale in the velocity field is at the same scale as where the magnetic field peaks.
Often when the spatial scales  of growing fields are comparable to the fluctuations 
use of the term "small scale dynamo"  is employed so some further clarification is warranted.
In the fiducial "small scale" or "fluctuation dynamo" commonly studied  in turbulently (non-helically) forced periodic boxes, 
the field grows independent  of a mean field.
In the kinematic phase for such dynamos
the turbulent stretching action of the flow counters diffusion leading to spatially self-similar growth
of the magnetic field on scales much smaller than the forcing scale \citep{kaz68, KA92, KS97}.
However upon nonlinear saturation, the field can become more coherent, 
with the peak coming closer to the forcing scale \citep{HBD04,BS13}.
This contrasts  the MRI dynamo that we have studied here
which first grows fields on the scales associated with the maximum growth rate for the MRI, which are always 
large scales in the radial direction in a given domain as shown earlier from the vertically averaged fields.
Thus the large scale  fields are then followed by the small scale fields unlike the fluctuation dynamo just described.

The large scale (planar averaged) fields produced by the MRI dynamo eventually exhibit secular cycle periods of $\sim 10$ orbits,
evolving on long time scales compared to the orbit or shear time scale. 
In at least that sense, the MRI large scale dynamo resembles other  large scale dynamos with cycle periods, 
even though there may not be spatial scale separation between the fluctuations and mean fields.
In this respect the large scale MRI dynamo as we have self-consistently defined it (exponential growth of planar averaged fields) does not have 
as part of its definition, that the  fluctuations need to have gradient scales   much smaller than the scales on which mean quantities vary.  
This is important to keep in mind when comparing to other traditional contexts  of mean field dynamo for stars and galaxies.   
Overall we feel justified in the term "large scale dynamo" to describe the growth of planar
average fields in MRI simulations. We could also call the large scale dynamo a global dynamo 
with respect to simulation box size in the sense that the large scale fields are directly produced by global modes of the MRI.

\section{Conclusions}
\label{conclusion}
In summary, our five major findings are: 
1) that large-scale fields are generated and do persist in either types of horizontal or vertical averaging; 
2) through spectral analysis and space-time plots of averaged mean fields, we show the generation of large-scale fields 
even during the early growth phase of MRI for the first time, indicating dynamo action due to MRI even in absence of turbulence.
3) there is direct feedback from low wavenumber horizontal MRI modes to the initial vertical field, which can lead to sustenance of the mean vertical field
against turbulent diffusion in saturation regime.
4) from DNS, we compute the individual terms in the mean field induction equation using both types of averaging and identify their contribution to the generation of mean fields. 
5) the large scale fields arise first and then there is direct transfer from large to small scales on saturation.
Each is summarized further below.

\subsection{Vertical averaging and growth of large scale field in MRI growth phase}
We ran DNS of the MRI  with resolutions of $256\times1024\times256$ and $128\times512\times128$.
The  fields and stresses were converged as shown in \Figs{timeplot}{stress}.
Upon planar averaging in two different schemes ($x$-$y$ and  $y$-$z$ averaging), 
stronger large scale fields were found in the component along the direction of the shear flow. 
The $y$-$z$ averaging in particular, exhibits radially varying mean fields which compare well
with the global cylindrical simulations of \citet{EB16}.
We identified a distinctive "loop" feature
just before saturation of the $x$-$y$ averaged field (as can be seen in the left panel of \Fig{horxy}),
as indicative of the modal quasilinear nature of MRI growth phase.
Upon planar averaging  in the MRI growth phase,  the field depicted by logarithmic color scaling in \Fig{linGrow}, 
reveals  growth of coherent modal mean fields.

Finally, either horizontal or vertical
averaging results in large scale fields in our shearing box simulations and new lessons emerge by comparing these two types of averaging.
Vertical planar averaging reveal results consistent with  previous global cylindrical MRI model of vertically averaged field growth. 
Second, only in the vertical $y$-$z$ averaging, do we see direct feedback of MRI large scale modes on the initial vertical field, which can sustain the vertical field against turbulent diffusion.
Previous MRI shearing box studies focus on horizontal averaging, but  we find that vertical averaging 
yields conceptual fruit in part because the direction of velocity gradient ($x$) is left unaveraged. 

We also find that the basic conceptual importance of large scale fields for MRI unstable systems is compatible with implications of  \cite{NB14} and \cite{BN15}  from different calculations.

\subsection{Dynamo action in the absence of turbulence}
We analysed the MRI growth phase by evaluating  power spectra for each component of the field
along all of the wavenumbers ($\tilde{k}_x,\tilde{k}_y,\tilde{k}_z$) as shown in \Fig{spec}.
We find that due to the global nature of the instability, there are large scale modes
growing in $\tilde{k}_x, \tilde{k}_y$ in the first few wavenumbers.
The field  resides in a narrow range  of scales and not until saturation
does evidence for turbulence manifest itself.
Thus we find dynamo action in the early growth phase without turbulence, consistent with the single mode
analysis in \citet{EB16}. 

There is a correspondence between the modal structure of the mean field obtained from planar averaging in the MRI growth phase, 
with the dominant mode in the power spectrum.
The linear non-axisymmetric MRI has been well studied in previous work which shows 
field  growth at nonzero horizontal wavenumbers 
(see Fig.8 in \citet{Khalzov2006}) \citep{BH92}. Here we have shown that these low wavenumber modes
survive planar averaging.

\subsection{Feedback to vertical mean field}
The horizontal  low wavenumber MRI modes are seen to grow in all vector components of the field, 
which includes feeding back also to the vertical mean field.
By subtracting the initial field from the evolving total vertical field and then vertically averaging, we 
find a finite mean field growing as shown in \Fig{bzcomp}. 
This feedback from MRI unstable low wavenumber modes to the vertical field has now been identified here,
and is important for countering turbulent diffusion in the nonlinear saturated regime.
The importance of this feedback obtains also from the fact that in the absence of any
such large scale dynamo action, the arising MRI turbulence would destroy the original vertical field
and can then shutdown the MRI.

We have not explored the saturation of the MRI large scale dynamo in this paper
but it is a topic of our further investigation.
We have not varied the geometry of the initial field condition in our simulations,
other than to identify similar behavior between
zero net flux and a net flux simulations.
But given the global nature of MRI, we would expect  similar behaviour
for a range of initial field geometries.   

\subsection{EMF responsible for mean field growth in MRI dynamo}
In \Figs{mrimft1}{mrimft2} we show the contribution of individual terms in the mean field equations
to field growth. We find that in the MRI growth regime, the time derivative of the mean field 
is well matched by using only the spatial derivative of EMF terms in vertical averaging. 
Whereas in the case of horizontal averaging, 
the shear term ($S\overline{B}_x$) for $\overline{B}_y(z)$) also contributes.
But in the nonlinear saturation regime, for  both cases, the EMF term contributions dominate. 
We also find that the contribution from the stretching and advection terms involving only mean fields are always negligible.   
The results of vertical averaging compare well with the global simulations of EB16, 
where the large scale fields arise entirely from the EMF. 

\subsection{Direct transfer of energy from large scales to small scales}
\Figs{linGrow}{spec} show that for the MRI dynamo,  large scale modes of the large scale fields
grow first in the MRI growth phase followed by 
small scale fields, as the power spectra broaden towards saturation. 
The planar averaged fields themselves develop smaller gradient scales 
suggesting the presence of turbulence in \Fig{horxy}.
Since the field is not yet turbulent in the early MRI growth phase, a turbulent diffusion tensor is not expected to be present.
Thus for the growth phase, this contrasts the dynamo mechanism proposed by \citet{SB15}.

\subsection{Future work}
Among the questions that remain include 1) a robust
numerical testing of  how growth proceeds at much lower initial seed fields.
In particular, numerical limitations prevent starting with too low a seed field strength
so it was difficult to assess whether the exponentially growing modes in the vertical field
for vertical averaging would contribute significantly to net total amplification of the initial vertical field
by the time of saturation.
2) How do the  small scale fields emerge (via mode-mode coupling) and lead to saturation?
3) What is the best model for the ensuing turbulent diffusion that balances growth in this stage?
These latter two questions are related to the evolution of the EMF from
the MRI growth to the nonlinear regime, which we have not explored in this 
paper. An important goal is to understand what determines the saturation
amplitude of both the large scale mean field and the fluctuations, and their
connection to transport stresses.
The nonlinear nature of the dynamo  in the saturated regime 
does not necessarily preclude a mean field formulation but 
the resulting transport coefficients in the EMF are expected to be evolving functions of the field itself.
The extent to which magnetic  helicity evolution plays a role in 
the large scale dynamo saturation is also an open question.
All of this has implications for the steady state amplitude of the field and transport stresses,
and all are fruitful topics for  further investigation.

\section*{Acknowledgments}

We acknowledge useful discussions with K. Subramanian and  F.  Nauman.
PB and FE acknowledge grant support from DOE, DE-SC0012467.
EB acknowledges  support from  grants HST-AR-13916.002, and NSF AST1515648.
The computing resources were provided by Princeton Institute of Computational
Science (PICSciE).

\bibliographystyle{mn2e}
\bibliography{mrilsd}

\label{lastpage}

\end{document}